\documentclass[english,letterpaper,aps,prl,twocolumn]{revtex4-2}
\usepackage{lmodern}
\usepackage[T1]{fontenc}
\usepackage[latin1]{inputenc}
\usepackage{amstext}
\usepackage{amssymb}
\usepackage{graphicx}
\usepackage{physics}
\makeatletter
\@ifundefined{textcolor}{}
{%
 \definecolor{BLACK}{gray}{0}
 \definecolor{WHITE}{gray}{1}
 \definecolor{RED}{rgb}{1,0,0}
 \definecolor{GREEN}{rgb}{0,1,0}
 \definecolor{BLUE}{rgb}{0,0,1}
 \definecolor{CYAN}{cmyk}{1,0,0,0}
 \definecolor{MAGENTA}{cmyk}{0,1,0,0}
 \definecolor{YELLOW}{cmyk}{0,0,1,0}
}

\usepackage{lmodern}

\makeatletter

\usepackage{lmodern}

\makeatletter

\usepackage{lmodern}

\makeatletter

\makeatletter

\usepackage{epsfig}

\usepackage{dcolumn}

\usepackage{bm}

\usepackage{bbm}

\usepackage{wasysym}

\usepackage{marvosym}

\usepackage{epstopdf}

\usepackage{subfloat}
\usepackage{color}
\usepackage{subfigure}

\usepackage{blindtext}

\newlength{\textwidthm}

\makeatother

\makeatother

\usepackage{tikz}

\makeatother

\makeatother

\makeatother

\usepackage{babel}
\begin{document}
\title{Disorder free many-body localization transition in two quasiperiodically
coupled Heisenberg spin chains}
\author{K.G.S.H. Gunawardana$^{1}$ and Bruno Uchoa$^{2}$ $^{*}$}
\affiliation{$^{1}$Department of Engineering Technology, Faculty of Technology,
University of Ruhuna, Matara, Sri Lanka}
\affiliation{$^{2}$Department of Physics and Astronomy, University of Oklahoma,
Norman, OK 73069, USA}
\email{harshadewa@etec.ruh.ac.lk, uchoa@ou.edu }

\date{\today}
\begin{abstract}
Disorder free many-body localization (MBL) can occur in interacting
systems that can dynamically generate their own disorder. We address
the thermal-MBL phase transition of two isotropic Heisenberg spin
chains that are quasi-periodically coupled to each other. The spin
chains are incommensurate and are coupled through a short range exchange
interaction of the $XXZ$ type that decays exponentially with the
distance. Using exact diagonalization, matrix product states and density
matrix renormalization group, we calculate the time evolution of the
entanglement entropy at long times and extract the inverse participation
ratio in the thermodynamic limit. We show that this system has a robust
MBL phase. We establish the phase diagram with the onset of MBL as
a function of the interchain exchange coupling and of the incommensuration
between the spin chains. The Ising limit of the interchain interaction
optimizes the stability of the MBL phase over a broad range of incommensurations
above a given critical exchange coupling. Incorporation of interchain
spin flips significantly enhances entanglement between the spin chains
and produces delocalization, favoring a pre-thermal phase whose entanglement
entropy grows logarithmically with time. 
\end{abstract}
\maketitle

\section{Introduction}

Many-body localization (MBL) describes a dynamical phase of an interacting
quantum system that can not reach thermal equilibrium in the thermodynamic
limit \cite{Anderson,Nandkishore,Rigol1,Abanin,Abanin2,Imbrie,Huse,Altman,Huse2,Basko}.
The growth of entanglement with time within an isolated system is
inhibited in the MBL phase, resulting in nonergodic time evolution
and area law scaling of the entanglement entropy. A thermal phase
in contrast follows ergordic time evolution, developing full entanglement
in the Hilbert space and volume law scaling of entanglement entropy.
Thus, a hallmark of the MBL phase is the onset of very slow dynamics
that preserves information of the initial quantum state \cite{Badarson,Singh,Serbin,Luitz,Schreiber}.
MBL states have been experimentally observed in optical lattices with
cold atoms systems \cite{Schreiber,Lukin}, where the entanglement
entropy can be directly measured \cite{Islam}, and also in circuits
with superconducting qubits \cite{Houck,Roushan}. Those states are
of technological importance in the development of quantum memory \cite{Throckmorton,Guo,Zhang,Shan}
and also of fundamental interest to subjects ranging from quantum
information, time crystals and quantum thermalization in closed systems
\cite{Deutsch,Srednicki,Rigol,D'alessio,Devakul}.

Disorder and interactions are identified as key controlling parameters
driving a thermal system towards a MBL phase. In the absence of interactions,
a quantum system subjected to arbitrarily weak disorder potential
would be Anderson localized \cite{Anderson,Billy,Yamilov} in 1D.
Many-body localization occurs in one dimensional interacting systems
in the presence of externally applied random disorder fields \cite{Basko,Badarson,Singh,Serbin,Luitz,Oganesyan,Pal,Kjall,Chandran,Vu}.
A system can exhibit thermal-MBL phase transition when the magnitude
of the disorder strength ($h$) is greater than a critical value,
$h>h_{c}$. Usually, this transition occurs through a marginally localized
intermediate regime, which may depend on the system size, disorder
strength and the interactions \cite{Luitz,Vu,Morningstar}.

MBL has been theoretically proposed in the presence of quasiperiodic
static potentials described by the Aubry-Andre model in 1D \cite{Schreiber,Thouless,Iyer,Vidal,Hauschild}.
The onset of thermal-MBL phase transition in this model has been found
to be at the critical value $h_{c}=2$ (in units of the ``kinetic''
energy) and followed by a broad marginally localized precursor to
the MBL regime \cite{Vu,Thouless,Iyer}. Recent numerical studies of two-leg
ladder model provide signatures of the thermal-MBL transition for
both random disorder and the Aubry-Andre model under an externally
applied critical field strength in the range $8<h_{c}<10$ \cite{Hauschild,Baygan}. 

Disorder free MBL arises in systems that can dynamically generate
their own disorder in the absence of externally applied fields. Proposals
in 1D lattices include of out-of-equilibrium bosons \cite{Carleo}
or spins \cite{Hickey}, or families of models with fermions effectively
coupled to spins \cite{Smith,Smith2}. A two-leg ladder compass model,
with discrete translational symmetry and imposed topological constraints
on the Hilbert space due to conservation laws, identified a pre-thermal
phase with logarithmic growth of the entanglement entropy in time
\cite{Hart}. It has been shown that the presence of a linearly varying potential in spin chains may disentangle the Hilbert space in discrete sectors, resulting in non-ergordic MBL-like dynamics \cite{Guo, Evert, Schulz, Ling, Taylor, YaoR,Doggen}.   In another proposal, two coupled fermionic chains with
full translational symmetry, each chain having a different species
with either heavy or light masses, were found to have early time evolution
indications of MBL \cite{Schiulaz,Grover,Schiulaz2,Papic2,Yao}. Translationally
invariant systems, nevertheless, show strong finite-size effects and
are expected to delocalize in the thermodynamic limit at long times
\cite{Papic2,Yao}. Moreover, implementation of some of the other
proposals requires the preparation of specifically ordered states
in finely tuned Hamiltonians.

\begin{figure}[t]
\includegraphics[scale=0.34]{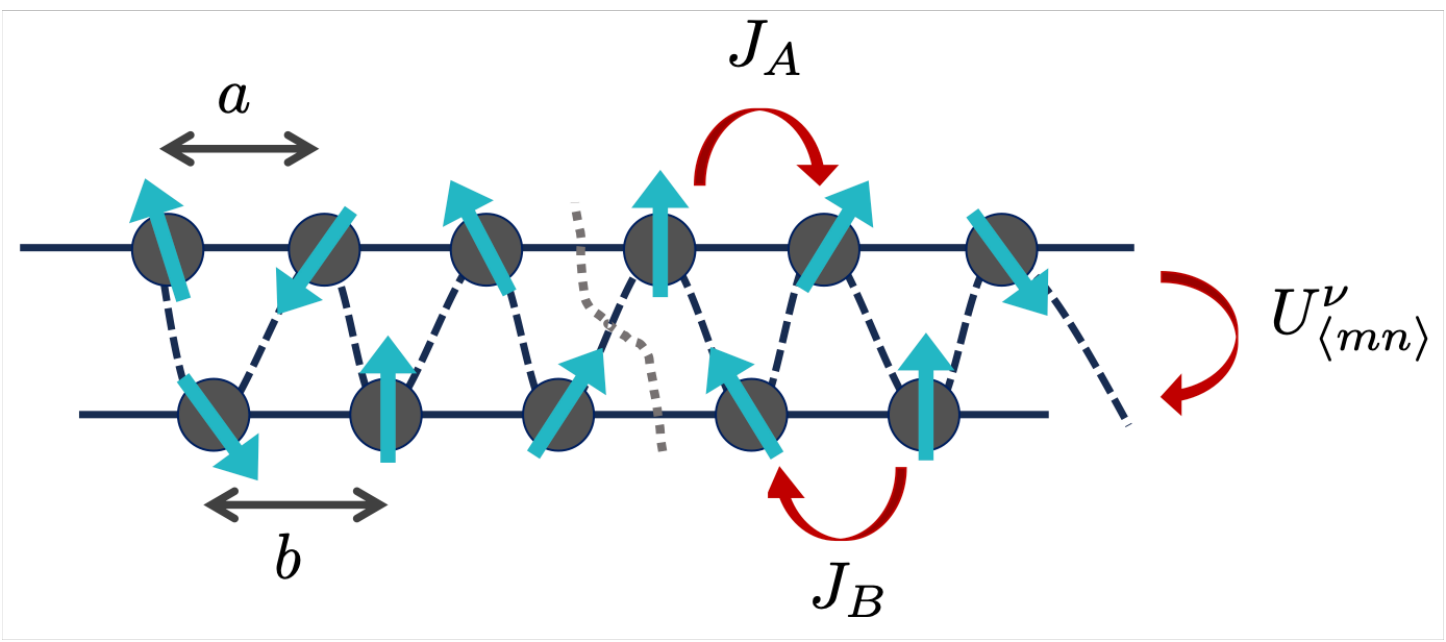}
\caption{\label{fig:model} Schematic of the two isotropic Heisenberg spin
chains having lattice parameters $a$ and $b$. The isotropic spin
interaction within each chain is described by $J_{A}$ and $J_{B}$.
The short range NN exchange coupling between chains at direction $\nu$,
$U_{\langle nm\rangle}^{\nu}$, is guided by the black dashed line.
The dotted curvy line represents a bipartite cut employed to calculate
the time evolution of the entanglement entropy between left and right
parts of the system.}
\end{figure}

In this work we address the question of whether the quasiperiodic coupling
between two isotropic spin chains, each one with discrete translational
symmetry, can produce a robust MBL phase. We show that the answer is
affirmative. We propose a model consisting of two isotropic spin chains
coupled to each other by an anisotropic short-range exchange interaction
of the $XXZ$ type that decays exponentially with distance. The two
spin chains have different incommensurate lattice parameters $a$
and $b$. The ratio between lattice parameters is irrational,
\begin{equation}
\frac{b-a}{a}\equiv\gamma\delta,\label{eq:incom_lattice}
\end{equation}
where $\delta>0$ is a real number and $\gamma$ is some irrational
number whose value is chosen to be $\gamma=(1-\pi^{2}/10)$, with
$0\leq\gamma\delta<1$. This construction results in two coupled spin
chains with incommensurate lattice constants. Because their exchange
coupling decays exponentially with the distance between spin sites,
the two chains are quasi-periodically coupled to each other, as indicated
Fig. 1. 

In the Ising limit of the $XXZ$ exchange between the chains, we show
that this system enters a MBL phase above a critical value of the
quasiperiodic interchain exchange coupling. Such critical value is
strongly dependent on the incommensuration. We provide numerical evidence
that the MBL phase is optimized for $0.176<\gamma\delta<0.712$ and
is suppressed in the commensurate limit $\gamma\delta\to0$. In the
optimal regime, the MBL phase emerges when the interchain exchange
coupling in the $z$ spin axis $U_{0}^{z}/J>9$, with $J$ the isotropic
intrachain Heisenberg exchange coupling, while a pre-thermal phase
appears between $6<U_{0}^{z}/J<9$. In the latter, the entanglement
entropy grows logarithmically with time. Below $U_{0}^{z}/J<6$ the
system is in the thermal phase for most incommensuration values. 

We find that restoration of the interchain exchange coupling in the
$x$ and $y$ spin directions, $U_{0}^{xy}$, significantly enhances
entanglement between the chains and produces delocalization. In the
isotropic limit $U_{0}^{xy}=U_{0}^{z}$, the system is always in the
thermal phase. For strong but finite anisotropy, the MBL phase is
stabilized at $U_{0}^{z}/J>30$ for $U_{0}^{x,y}/J=1$ near the optimal
incommensuration $\gamma\delta\approx0.391$. In this regime thermalization
occurs at $U_{0}^{z}/J<9$, with a broad pre-thermal region in between. 

The structure of the paper is as follows: in section II we describe
the Hamiltonian of the system and proceed to calculate the time evolution
of the bipartite entanglement entropy $S$ in section III. In the
Ising limit of the $XXZ$ interchain exchange, we show that the entanglement
entropy follows a transition from volume law to area law scaling at
finite incommensuration, as the exchange coupling between the chains
is increased. Next, using exact diagonalization at zero magnetization,
we calculate the averaged inverse participation ratio (IPR) for finite
system sizes as a function of the incommensuration and the exchange
coupling. We extrapolate the IPR to the thermodynamic limit and construct
the phase diagram separating the thermal and MBL phases. To check
for the stability of the MBL phase in the Ising limit of the $XXZ$
exchange between the chains, we restore the interchain exchange interaction
along the $x$ and $y$ spin directions. We examine the time evolution
of the entanglement entropy to show that the MBL phase remains stable,
although at a much larger interchain critical coupling $U_{0}^{z}$.
We also calculate the IPR of the ground state for very large system
sizes using density matrix renormalization group (DMRG) to gain insight
in the behavior of the system in the $\gamma\delta\to0$ limit. Finally,
in section IV we present our conclusions. 

\section{Coupled spin chains model}
\begin{center}
\begin{figure*}[ht]
\includegraphics[scale=0.28]{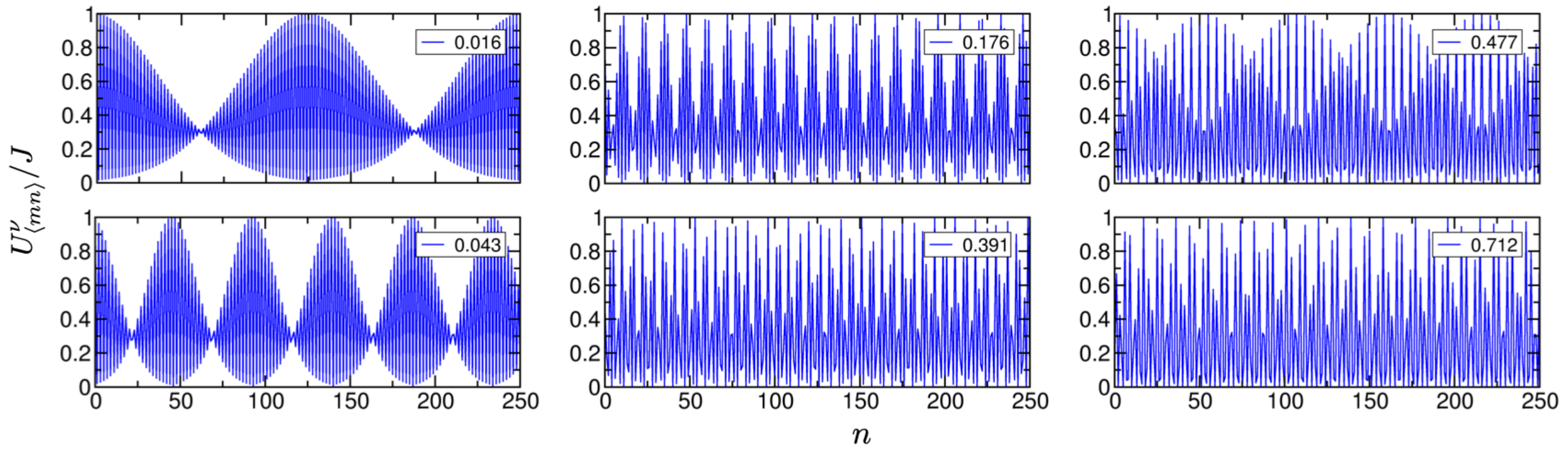}
\caption{\label{fig:pot} Evolution of the inter-chain interaction profile
for NN sites, $U_{\langle nm\rangle}^{\nu}/J$ versus the site position
$n$ in the spin ladder for different values of incommensuration $\gamma\delta$,
with $\gamma=(1-\pi^{2}/10)$ an irrational number. $n$ sites below
to chain $A$ and $m$ sites to chain $B$. The legends represent
the approximate value of $\gamma\delta$ for each potential profile.
$\delta=e^{0.2}, e^{1.2}, e^{2.6}, e^{3.4}, e^{3.6},$ and $ e^{4}$ for $\gamma\delta\approx0.016,0.043,0.176,0.391,0.477$
and $0.712$, respectively.}
\end{figure*}
\par\end{center}

We consider two isotropic spin-$\frac{1}{2}$ Heisenberg chains with
Hamiltonian
\begin{equation}
\mathcal{H}_{\alpha}=J_{\alpha}\sum_{i=1}^{N_{\alpha}}\mathbf{S}_{i}\cdot\mathbf{S}_{i+1},\label{eq:Heisenberg}
\end{equation}
where $\alpha=A,B$ labels each chain, and $J_{\alpha}>0$ is the
intrachain nearest neighbors (NN) exchange coupling. $N_{\alpha}$
is the number of spins on chain $\alpha$ and $\mathbf{S}=(S^{x},S^{y},S^{z})$
is the spin operator $S^{\nu}=\frac{\hbar}{2}\sigma^{\nu}$, with
$\nu=x,y,z$ labeling the standard Pauli matrices. The spin chains
are coupled to each other through the $XXZ$ exchange
\begin{equation}
\mathcal{H}_{AB}=\sum_{\nu=x,y,z}\sum_{n\in A,m\in B}U_{nm}^{\nu}S_{n}^{\nu}S_{m}^{\nu},\label{eq:HAB}
\end{equation}
where $U_{nm}^{x}=U_{nm}^{y}\equiv U_{nm}^{xy}$ and $U_{nm}^{z}$
are the interchain exchange couplings for spins oriented in the $x,y,z$
directions. The interchain exchange decays exponentially with the
distance between sites,
\begin{equation}
U_{nm}^{\nu}=U_{0}^{\nu}\text{e}^{\rho\left(1-\sqrt{1+r_{nm}^{2}}\right)},\label{eq:U}
\end{equation}
where $r_{nm}=|R_{n}^{A}-R_{m}^{B}|/d$ is the horizontal distance
between sites normalized by the distance between the two chains $d$,
and $\rho$ sets the range of the interaction. $R_{n}^{\nu}$ is the
position of the spins along the chains,
\begin{equation}
R_{n}^{A}=an-r_{0}^{A},\qquad R_{m}^{B}=bm-r_{0}^{B}.\label{eq:lattice}
\end{equation}
with $a$ and $b$ the lattice parameters of spin chains $A$ and
$B$ respectively, and $r_{0}^{\alpha}$ is the origin of each chain. 

The primary focus of this work is to investigate the effect of interchain
exchange coupling in the onset of the MBL transition. We finely tune
$J_{A}=J_{B}=J$ so that in the limit $U_{0}^{\nu}\rightarrow0$ the
system decouples into two identical isotropic Heisenberg spin chains,
which are ergodic. The exchange coupling between chains $U_{\langle nm\rangle}^{\nu}$
given in Eq. (\ref{eq:HAB}) is truncated to the two NN spin sites,
as shown in Fig. 1. The solid lines connecting spin sites in each
chain represent the isotropic interactions between spins in each chain.
The dashed line running between spin sites is a guide to the eye representing
the short range exchange coupling between chains. Thus, each spin
in a chain couples through $J$ with two NNs in the same chain and
through $U_{\langle nm\rangle}^{\nu}$ with upto two nearest spins
in the opposite chain. We set $\rho=10$ in Eq. (\ref{eq:U}). Our
conclusions do not depend on the choice of $\rho$, which will at
most rescale the localization length at finite system sizes, but
not in the thermodynamic limit. 

We consider the regime where the incommensuration is in the range
$0\le\gamma\delta<1$, in which the lattice parameters satisfy $b\in[a,2a]$.
For large incommensurations $\gamma\delta\gg1$, the linear bond density
between chains is reduced, as spin chain $B$ becomes sparse, and
the two spin chains effectively decouple. The evolution of the profile
of the interchain interaction between NN spins $U_{\langle nm\rangle}^{\nu}$
with the incommensuration is shown in Fig. 2. 

\begin{center}
\begin{figure*}[ht]
\includegraphics[scale=0.4]{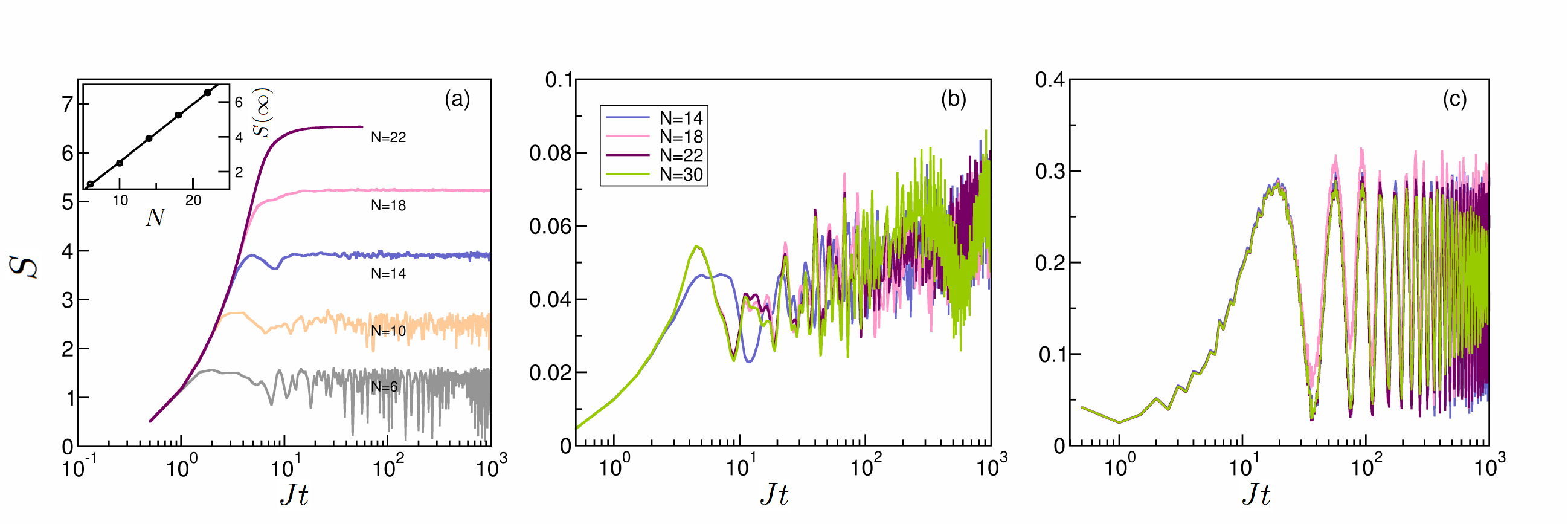}
\caption{\label{fig:1Sp} Time evolution of the bipartite entanglement entropy
$S(t)$ following a quantum quench from an initial product state for
different system sizes. (a) Thermal phase at $\delta=1.0$ ($\gamma\delta\sim0.013$),
$U^{z}_{0}/J=1$ and $U^{xy}_{0}=0$ for $N=6,$10, 14, 18 and 22.
The entropy saturates due to the finite size effects. The saturated
entropy follows the volume law of entanglement. Inset: Plot of the
saturated entropy $S(\infty)$ vs. number of ladder spins $N$ in
the thermal phase. The slope of the straight line is 0.3305. MBL phase
at $\delta=15.0$ ($\gamma\delta\sim0.196$), $U^{z}_{0}/J=25$ and
$U^{xy}_{0}=0$ for (b) $r_{0}^{A}=Na/2+0.5$ and $r_{0}^{B}=Nb/2+0.6$
and (c) $r_{0}^{A}=Na/2$ and $r_{0}^{B}=Nb/2+0.3$. Different curves
correspond to different system sizes. The relative placement of the
two chains is adjusted to keep bond strength at the bipartite cut
independent of the system size. In either case, the curves collapse
on top of each other, consistently with area law entanglement (see
text).}
\end{figure*}
\par\end{center}

\section{Methods and Results}

The many-body quantum states of the total Hamiltonian
\begin{equation}
\mathcal{H}=\mathcal{H}_{A}+\mathcal{H}_{B}+\mathcal{H}_{AB}\label{eq:Htotal}
\end{equation}
can be described using $2^{N}$ basis vectors spanning the Hilbert
space, with $N=N_{A}+N_{B}$ the total number of sites in the spin
ladder. The basis vectors are constructed with product states $|s_{1}\rangle\otimes\cdots\otimes|s_{N}\rangle$,
where $|s_{i}\rangle=|\uparrow\rangle,|\downarrow\rangle$ are the
eigenstates of $S_{i}^{z}$ on site $i$. By convention, the product
states are arranged from left to right in ascending order as guided
by the dashed line in Fig. 1. We use the above ordered set of basis
vectors to calculate the matrix elements of $\mathcal{H}$ and construct
the matrix product state (MPS) in our numerical calculations.

We numerically calculate the time evolution of an initial product
state following a global quantum quench using the unitary transformation,
$|\psi(t)\rangle=e^{-i\mathcal{H}t}|\psi_{0}\rangle$. The initial
state is chosen so that the spins in each chain are arranged in a
Neel state. For instance, at low incommensuration ($\gamma\delta\approx0)$
the initial product state is given by $|\psi_{0}\rangle=|\uparrow,\uparrow,\downarrow,\downarrow,\cdots,\uparrow,\uparrow\rangle$.
We use MPS to represent the quantum many body system and study the
dynamics following the time-evolving block decimation (TEBD) method.
 In TEBD method, the time evolution operator $e^{-iH\tau}$is decomposed
into the product of locally interacting pair of spins using a second
order trotter decomposition and contracted with the MPS to obtain
the updated quantum state after time $\tau$ (see Appendix \ref{A1}). This process is repeated
$t/\tau$ time steps to obtain the quantum state of the system after
time $t$. The maximum time step used is $\tau=0.05J^{-1}$. The internal
bond dimension of the MPS can be truncated in low entangled systems
to improve computational efficiency. In this work, we use a weight
cutoff of $10^{-7}$ in the MPS of the thermal phase. In the MBL phase,
we used a MPS bond dimension of 50.

\subsection{Entanglement Entropy}

The spin ladder is bipartitioned with a vertical cut into two spin
ladders with $N/2$ of spins, as shown in Fig. 1. We study the development
of entanglement between the left and the right half of the system
with time. The reduced density matrix of the left half of the system
$(\rho_{L})$ is calculated by tracing out the quantum degrees of
freedom of right half. This quantity is a probabilistic measure of
the entanglement developed through $2^{N/2}$ bonds between the left
and the right halves of the system. Thus, the Von-Neumann bipartite
entanglement entropy is calculated as $S=-\text{tr}_{L}\left[\rho_{L}\ln(\rho_{L})\right]$.

\textbf{Volume law versus area law. }We investigate the role of the
exchange coupling between chains and the amount of incommensuration
in the MBL transition. We first turnoff the spin flip exchange interaction
between two chains by setting $U_{0}^{xy}=0$. Choosing the relative
position of the two spin chains to be $r_{0}^{A}=r_{0}^{B}=0$, we
show in Fig. \ref{fig:1Sp} that the bipartite entanglement entropy
$S$ follows a transition from volume law entanglement to area law
as a function of $U_{0}^{z}$ and $\gamma\delta$. Fig. \ref{fig:1Sp}a
shows the time evolution of $S$ for $\delta=1.0 $ ($\gamma\delta\approx0.013)$
and $U_{0}^{z}/J=1$ at different system sizes. As time elapses, $S$
initially grows quickly and eventually saturates close to the maximum
possible value of entanglement entropy for half system, $N\ln(2)/2\approx0.3466N$.
 The saturated entropy $S(\infty)$ scales linearly with the size
of the system $N$ (see inset of Fig. \ref{fig:1Sp}a) with a slope
of $0.3305$. Thus, the system is in the thermal phase and follows
a volume law of entanglement. Those results are independent of the
choices of $r_{0}^{A}$ and $r_{0}^{B}$.

At higher values of $U_{0}^{z}$ and incommensuration the system has
area law entanglement across the left and right partitions. In Fig.
\ref{fig:1Sp}b and c, we plot the time evolution of $S$ for $\delta=15.0$
($\gamma\delta\approx0.196$) and $U_{0}^{z}/J=25$ in four different
system sizes: $N=14$, $18$, $22$ and $30$. We adjust the relative
placement of the chains through $r_{0}^{A}$ and $r_{0}^{B}$ to ensure
that the bond strength $U_{\langle\bar{n}\bar{m}\rangle}^{z}$ at
the bipartite cut $(\bar{n}=N/2)$ is independent of the chain size.
Fig. \ref{fig:1Sp}b depicts the time evolution of the $S(t)$ curves for
$r_{0}^{A}=Na/2+0.5$ and $r_{0}^{B}=Nb/2+0.6$, where the bond strength
at the partition is $U_{\langle\bar{n}\bar{m}\rangle}^{z}=0.46$ for
all $N$. Fig. \ref{fig:1Sp}c depicts the behavior of $S(t)$ for
a different choice of relative placement of the chains, $r_{0}^{A}=Na/2$
and $r_{0}^{B}=Nb/2+0.3$, where the bond strength at the bipartite
cut is $U_{\langle\bar{n},\bar{m}\rangle}^{z}=1.2$. In both cases,
$S(t)$ is independent of $N$. The different curves for several chain
sizes collapse into a single curve. The system has thus area law entanglement
and is many-body localized up to the longest time scales $t\sim10^{3}J^{-1}$
observed in our simulations. 
\begin{center}
\begin{figure*}[ht]
\includegraphics[scale=0.27]{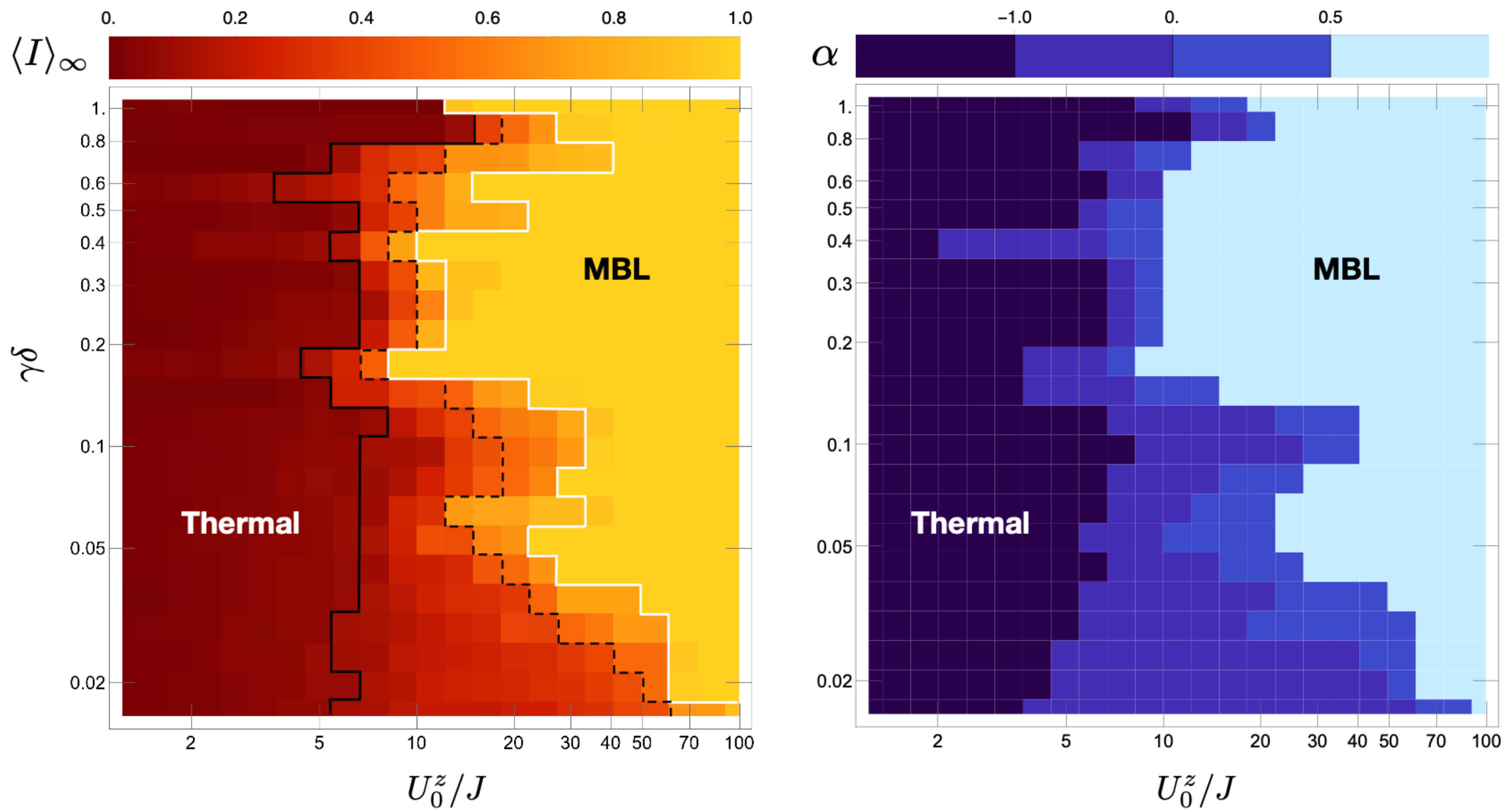}
\caption{\label{fig:qphase} Numerical phase diagram between thermal and MBL
phases. Left panel: color plot of the mean inverse participation ratio
extrapolated to the thermodynamic limit $\left\langle I\right\rangle _{\infty}$
versus incommensuration $\gamma\delta$ and strength of exchange coupling
between the chains $U_{0}^{z}/J$. Right panel: Color plot of the exponent
$\alpha$ extracted in the extrapolation of $\left\langle I\right\rangle $
to the thermodynamic limit versus $\gamma\delta$ and $U_{0}^{z}/J$.
The onset of the thermal-MBL phase separation (black solid line) is
drawn at $\left\langle I\right\rangle _{\infty}=0.1$. The thermal
region (dark red) to the left of the line corresponds approximately
to the region where scaling exponent $\alpha<-1.0$ in the right panel.
The black dashed line is drawn at $\left\langle I\right\rangle _{\infty}=0.5$.
The solid while line is at $\left\langle I\right\rangle _{\infty}=0.9$
and the region to its right in bright yellow is in the MBL phase.
This region approximately matches the light blue region on the right
panel, where $\alpha>0.5$. The intermediate region between the solid
black and solid white lines is marginally localized ($0.1<\langle I\rangle_{\infty}<0.9$).
This region corresponds to the range of scaling exponents $-1<\alpha<0.5$
in the panel on the right. }
\end{figure*}
\par\end{center}

\subsection{MBL phase diagram}

To quantify the MBL transition over a broader range of parameters
and extract the thermal-MBL phase diagram, we calculate the mean inverse
participation ratio (IPR) of the full energy spectrum at infinite
temperature using exact diagonalization. We impose that the system
has zero net magnetization and restrict the size of the Hilbert space
by picking only the $N!/\left(N/2)!\right)^{2}$ basis vectors that have the
same total number of $|\uparrow\rangle$ and $|\downarrow\rangle$
states. The Hamiltonian matrix of Eq.(\ref{eq:Htotal}) is diagonalized
to obtain the full eigenspectrum of the system. The IPR is calculated
through the average
\begin{equation}
\left\langle I\right\rangle =\frac{1}{D}\sum_{\lambda=1}^{D}\frac{4}{N}\sum_{i}^{N}\langle\phi_{\lambda}|S_{i}^{z}|\phi_{\lambda}\rangle^{2},
\end{equation}
where $|\phi_{\lambda}\rangle$ is the $\lambda$-th eigenvector and
$D$ is the number of eigenvectors in the system. For a maximally
localized (thermalized) phase, $\langle\phi_{\lambda}|S_{i}^{z}|\phi_{\lambda}\rangle=\pm\frac{1}{2}\,(0)$
at each spin site and hence $\left\langle I\right\rangle $ takes
the value $1$ ($0$). We calculate the IPR for each value of $U_{0}^{z}/J$
and incommensuration $\gamma\delta$ in five system sizes, $N=8,10,12,14,16 $,
and then extrapolate to the thermodynamic limit ($N\rightarrow\infty$),
$\left\langle I\right\rangle _{\infty}$. Following the procedure
described in ref. \cite{Vu}, we adopt the ansatz
\begin{equation}
\frac{\left\langle I\right\rangle }{1-\left\langle I\right\rangle }\propto N^{\alpha},\label{eq:IPR-norm}
\end{equation}
from which we extract the scaling exponent $\alpha$. In ideally thermalized
states, where the system is entirely delocalized and ergodic, $\alpha<-1$
with $\left\langle I\right\rangle _{\infty}\sim0$. On the other hand,
in the MBL phase, one expects $\alpha>0$ with $\left\langle I\right\rangle _{\infty}\sim1$. 

\textbf{Phase diagram.} The thermal-MBL phase diagram is drawn in
the left panel of Fig. (\ref{fig:qphase}), where we plot $\left\langle I\right\rangle _{\infty}$
against the interchain coupling $U_{0}^{z}$ and the incommensuration
$\gamma\delta$. The black solid line, drawn at $\left\langle I\right\rangle _{\infty}=0.1$
separates the thermal phase (dark red) from the marginally localized
and MBL phases (light red and bright yellow, respectively). The system
is in the thermal phase for all values of $U_{0}^{z}/J$ below the
the critical exchange coupling at the phase separation line (solid
black). This phase separation approximately corresponds to the boundary
of the region in the right panel of Fig. \ref{fig:qphase} where the
scaling exponent $\alpha<-1$. The system is in the thermal phase
when $U_{0}^{z}/J\lesssim$ $6$ for all values of $\gamma\delta<0.8$. 

The region where $U_{0}^{z}/J$ is larger than the critical value
set by the solid white line is identified as fully many-body localized
(yellow region), where $\left\langle I\right\rangle _{\infty}>0.9$.
It correlates with the region shown in light blue in the right panel,
where $\alpha>0.5$. The black dashed line is drawn at $\left\langle I\right\rangle _{\infty}=0.5$
and approximates the boundary where $\alpha=0$ in the right panel.
It is clear that the onset of the MBL phase has a strong dependence
on the incommensuration. In both the analysis of $\left\langle I\right\rangle _{\infty}$
and the scaling exponent $\alpha$, the optimal incommensuration for
the onset of MBL is in the range $0.176<\gamma\delta<0.712$, where
the critical exchange coupling is $U_{0}^{z}/J\sim10$. At low incommensurations,
$\gamma\delta\lesssim0.1$, the MBL phase is strongly suppressed.
In particular, in the limit $\gamma\delta\to0$, where the coupling
between the spin chains becomes periodic, we observe no MBL. In the
opposite limit, $\gamma\delta\to1$, the bond density between the
chains decreases as the lattice constant of one chain becomes nearly
twice as the constant of the other chain. In this regime the critical
coupling needed for MBL increases. 

The light red region in the left panel of Fig. (\ref{fig:qphase})
corresponds to an intermediate phase with $0.1<\left\langle I\right\rangle _{\infty}<0.5$
where the states are marginally localized. This region approximately
matches the mid region in the right panel, where $-1<\alpha<0$. The
width of this region is very broad at low incommensuration, reflecting
the suppression of MBL, but narrows down at $\gamma\delta\geq0.176$.
This appears to be due to the development of rapid oscillations in
the profile of the exchange interaction between the chains $U_{\langle nm\rangle}^{z}$
at $\gamma\delta\approx0.176$, as shown in Fig. 2. 

The extrapolation of $\left\langle I\right\rangle $ to the thermodynamic
limit in small system sizes is meaningful for interaction potentials
that oscillate rapidly compared to the system size. The extrapolated
results appear to be fairly accurate in the region $\gamma\delta>0.2$.
The extrapolation becomes less accurate in the opposite regime, at
low incommensuration. In any case, we note that this procedure correctly
captures the expected suppression of MBL in the low incommensuration
limit $\gamma\delta\to0$. 

\begin{center}
\begin{figure}[t]
\includegraphics[scale=0.30]{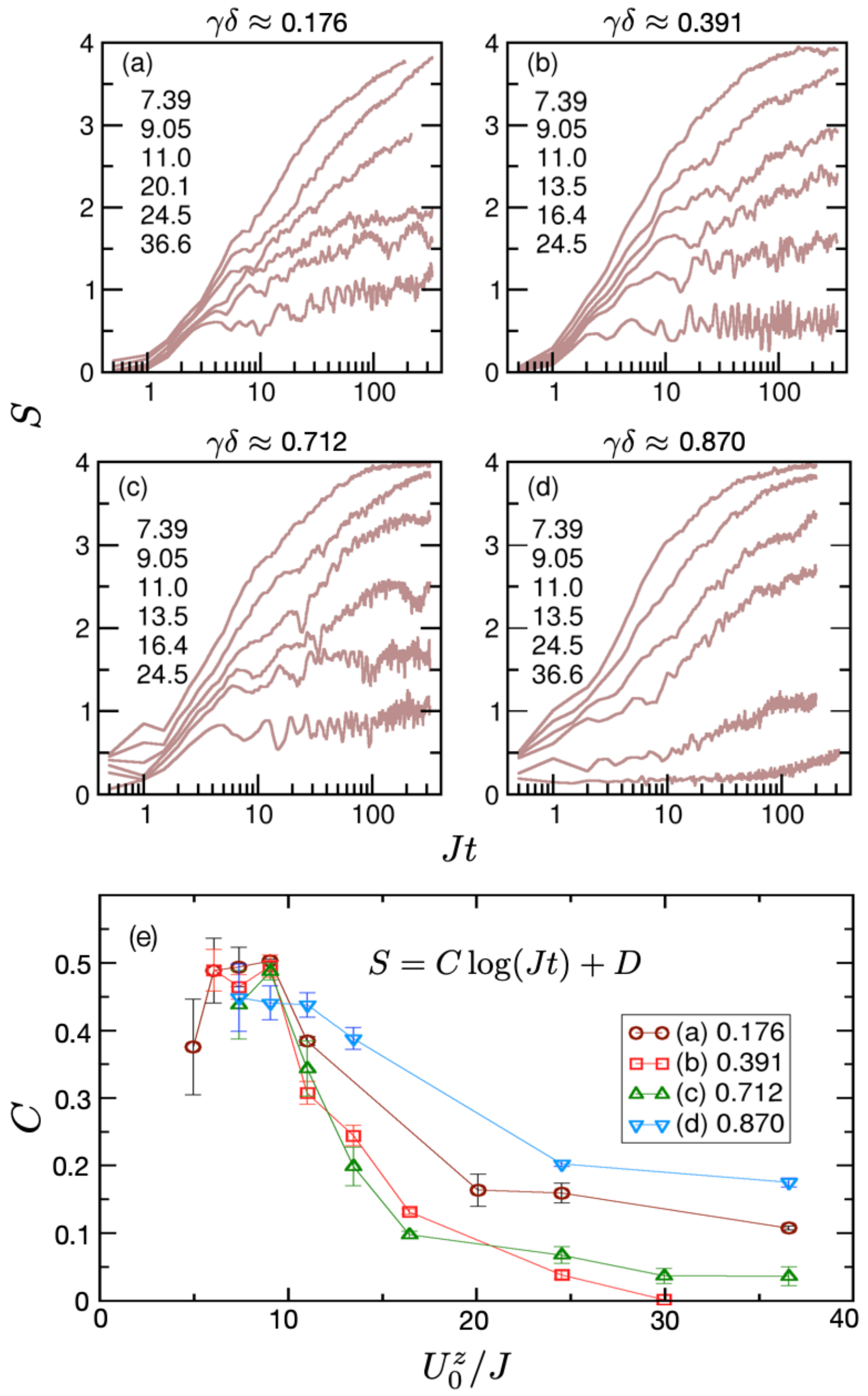}
\caption{\label{fig:S_Uxy} Time evolution of the bipartite entanglement entropy
$S$ at $U_{0}^{xy}/J=1$ for different incommensurations: (a) $\gamma\delta\approx0.176$,
(b) $\gamma\delta\approx0.391$, (c) $\gamma\delta\approx0.712$ and
(d) $\gamma\delta\approx0.870$. Top to bottom curves in each panel
correspond to ascending values of $U_{0}^{z}/J$ indicated on the
left of the curves. (e) Slope of the entanglement entropy $C$ extracted
from the logarithmic fit of the curves in panels (a) to (d). The system
enters in the MBL phase when $C=0$. }
\end{figure}
\par\end{center}

\subsection{Delocalization due to spin flips between chains}

We now turn on the exchange interaction between the spin chains $U_{0}^{xy}$,
which produces spin flips. Spin flips between chains map through the
standard Jordan-Wigner transformation into interchain hopping in the
fermionic language, which could lead to delocalization. We numerically
observe that this term drastically lowers $\langle I\rangle$ in the
finite size systems we simulated ($N\leq18$). Observation of MBL
through IPR over the whole energy spectrum would hence require much
larger system sizes in order to properly extrapolate the data to the
thermodynamic limit. This can be challenging given the exponential
growth in computational cost in exact diagonalization methods. To
gain insight, we resort to calculate the time evolution of the entanglement
entropy $S(t)$ through MPS. Even though MPS is relatively efficient,
interchain spin flip processes considerably increase the entanglement
of the states, requiring a much larger bond dimension for the MPS
compared to the $U_{0}^{xy}=0$ case. 

In the panels a$-$d of Fig. 5 we show the time evolution of the bipartite
entanglement entropy calculated from the initial state $|\uparrow,\downarrow,\uparrow,\downarrow\dots\rangle$ for $U_{0}^{xy}/J=1$ at four different
incommensurations, $\delta=  e^{2.6}$ for $N=18$ and  $e^{3.4}, e^{4.0},$ and $e^{4.2}$  for $N=14$ ($\gamma\delta\approx0.176,0.391,0.712\text{ and }0.870$,
respectively). 
Each panel shows six curves ordered from top to bottom
with increasing values of $U_{0}^{z}/J$ ranging from $7$ to $36$. 
We observe logarithmic growth of $S$ at long times before reaching
the saturation, a characteristic signature of a precursor to the MBL
phase in the parameter space. At sufficiently long times, this phase
is expected to thermalize.

In order to characterize the MBL transition, we fit the curves with
the form $S(t)=C\text{log}(Jt)+D$ and extract the slope $C$. In
panel 5e we plot the slope of the curves as a function of $U_{0}^{z}$.
We average over two different initial product states, $|\uparrow,\downarrow,\uparrow,\downarrow\dots\rangle$ and $|\uparrow,\uparrow,\downarrow,\downarrow\dots\rangle$,  and different initial
times. The values of $C$ for different system sizes are scaled to $N=14$. 
The behavior of $C$ with $U_{0}^{z}$ shows a broad peak followed
by a monotonic decrease with increasing $U_{0}^{z}$ starting at $U_{0}^{z}/J=9.2$
for all incommensurations inside the previously identified optimal
range $0.176<\gamma\delta<0.712$. The broad peak corresponds to a
thermal phase that saturates early due to finite size effects. The
point at which the slope starts to decrease monotonically with increasing
$U_{0}^{z}$ can be identified as the onset of the thermal-MBL transition,
with marginally localized states. The rate of decrease of the slope
$C$ varies with the incommensuration. The system enters in the MBL
phase only when $C\to0$. We note that this is the case at $\gamma\delta\approx0.391$
for $U_{0}^{z}/J\approx30$ where $C=0.001$. The slope nevertheless
decreases much slower for other incommensuration values, indicating
the broadening in size of the marginally localized region as the boundary
to the MBL phase retreats. This picture is qualitatively consistent
with an overall shift of the MBL phase separation line (white) to
the right in Fig. 4, combined with the emergence of a narrower range
of optimal incommensuration for the MBL phase that is centered around
$\gamma\delta\approx0.391$. 

\begin{center}
\begin{figure}[t]
\includegraphics[scale=0.17]{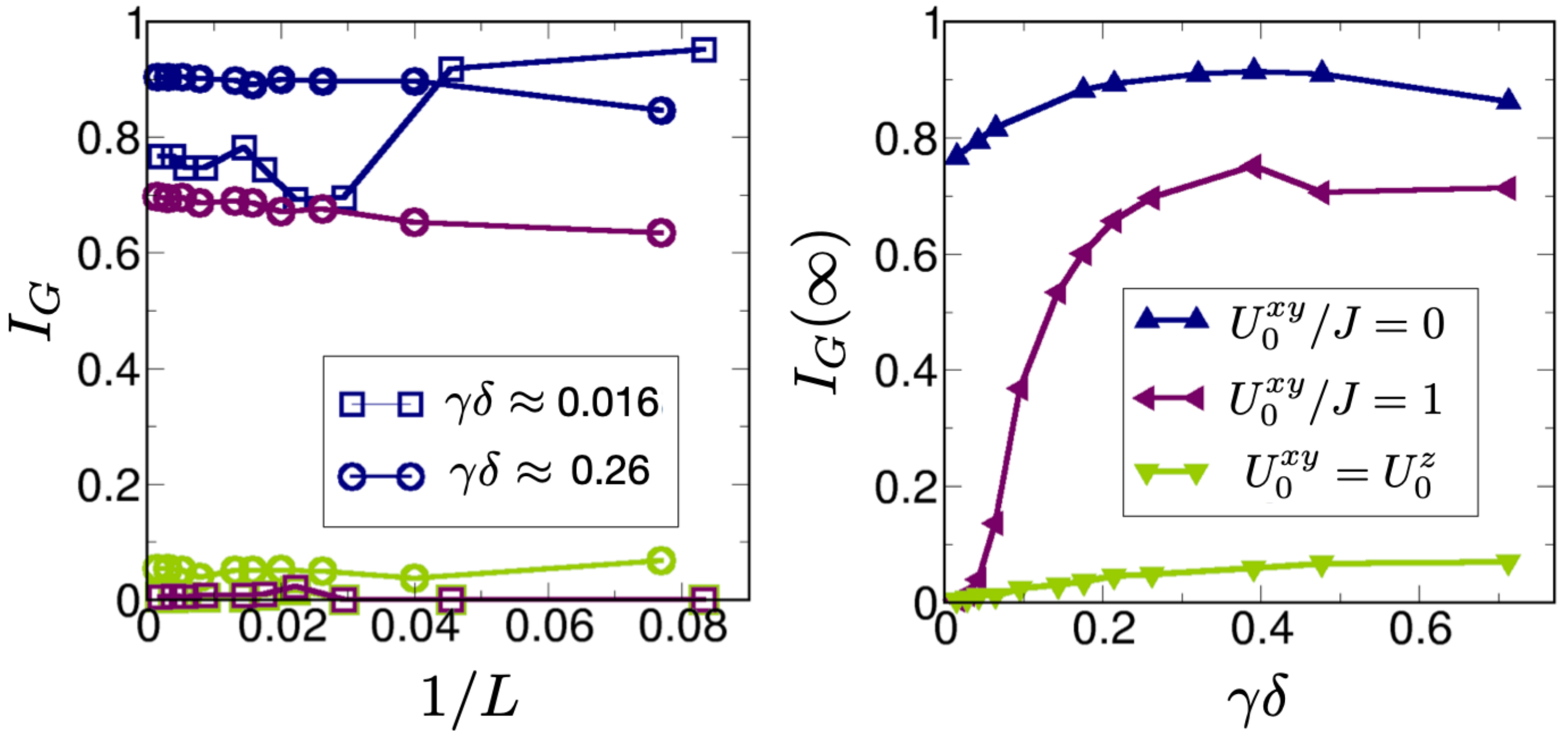}
\caption{\label{fig:deloc} DMRG calculation of the ground state inverse participation
ratio $I_{G}$ for $U_{0}^{z}/J=7.39$ and $U_{0}^{xy}/J=0,$ (blue
curves) $U_{0}^{xy}/J=1$, (purple) and $U_{0}^{xy}=U_{0}^{z}$ (green).
(a) $I_{G}$ versus inverse of the system size $1/L$ for $\gamma\delta\approx0.016$
(squares) and $\gamma\delta\approx0.26$ (circles). (b) Thermodynamic
limit of the ground state inverse participation ratio $I_{G}(\infty)$
for $L\to\infty$ versus incommensuration $\gamma\delta$.}
\end{figure}
\par\end{center}

To develop more insight on the effect of delocalization in the limit
of $\gamma\delta\rightarrow0$, we use DMRG to calculate the inverse
participation ratio of the ground state $I_{G}$ \cite{Schollok}. This quantity is
calculated for system sizes ranging from $N=24$ to $1100$, with
$L\sim\frac{N}{2}$. DMRG results are useful predictors for thermal phases. Even though the presence of localization
in the ground state does not inform about the behavior of the system
at infinite temperature, where it can delocalize, delocalization in
the ground state can conclusively rule out the emergence of MBL in
the thermodynamic limit. 

The plot in the left panel of Fig. \ref{fig:deloc} shows the variation
of $I_{G}$ with the inverse of the size of the chains $1/L$ at $U_{0}^{z}/J=7.39$
for $U_{0}^{xy}/J=0,1$ and $U_{0}^{xy}=U_{0}^{z}$ (blue, purple and
green curves respectively). The thermodynamic limit of $I_{G}$ at
$L\to\infty$ {[}$I_{G}(\infty)]$ for $U_{0}^{xy}/J=0$ (blue curves)
indicates localization of the ground state irrespective of the incommensuration
($\gamma\delta\approx0.016$ and $0.26$). This behavior is consistent
with the fact that the ground state of the isotropic antiferromagnetic
Heisenberg model for a single spin chain has gapped spinon excitations
\cite{PWAnderson, Faddeev}. Those gapped excitations remain stable in
the presence of an Ising exchange coupling with another chain ($U_{0}^{xy}=0$)
in the $\gamma\delta\to0$ limit. This is confirmed on the right panel
of Fig. 6, where we plot $I_{G}(\infty)$ as a function of the incommensuration
$\gamma\delta$. The blue curve shows that $I_{G}(\infty)\approx0.79$
at $\gamma\delta\approx0$. For the chosen set of couplings, the phase
diagram in Fig. 4 reveals that the system will eventually delocalize
in the infinite temperature regime through a pre-thermal phase. 

For finite anisotropy in the $XXZ$ exchange between the chains, we
observe a significant reduction of $I_{G}(0)$ with increasing $U_{0}^{xy}/J$.
For $U_{0}^{z}/J=7.39$ and $U_{0}^{xy}/J=1$ (purple solid line) $I_{G}(\infty)$
peaks at $\gamma\delta\approx0.391$, where it has a kink. This is
qualitatively consistent with MPS results for the time evolution of
the entanglement entropy shown in Fig. 5, which optimizes MBL at the
same incommensuration. At smaller values of $\gamma\delta$, $I_{G}(\infty)$
decreases rapidly and goes to zero in the $\gamma\delta\to0$ limit.
This suggests that the whole energy spectrum is delocalized in the commensurate
limit. In the isotropic limit of the $XXZ$ exchange between the chains,
$U_{0}^{xy}=U_{0}^{z}$, $I_{G}(\infty)\ll1$ for all incommensurations
(green curves in Fig. 6), consistently with a thermal phase. 

\section{Conclusion}

We showed that robust MBL emerges from two quasiperiodically coupled
Heisenberg spin chains. Using a combination of different numerical
methods, we derived the thermal-MBL phase diagram of this problem
as a function of the interchain exchange coupling and the incommensuration.
We show that the MBL phase is optimal in the Ising limit of the XXZ
interchain exchange interaction, over a whole range of incommensurations $0.176<\gamma\delta <0.712$. 
Spin flip processes between chains produce a significant amount of entanglement and favor a pre-thermal phase. 
MBL is generically present at finite incommensuration above a critical exchange coupling $U_0^z$ in the anisotropic regime $U_0^{xy}/U_0^z \ll 1$, and is 
entirely suppressed in the isotropic limit of the XXZ exchange interaction. 
This proposal does not require finely tuned Hamiltonians and could
be implemented in spin chains constructed in the absence of externally
applied potentials. 

We thank J. Knolle and A. Auerbach for insightful discussions. BU
was supported by NSF (US) DMR-2024864. KGSH acknowledges UOR for support. 

\appendix

\section{Numerical calculation of the time evolution of state $|\psi\rangle$ \label{A1}}

In this appendix, we provide numerical details of the time evolving block decimation (TEBD) method adopted  to describe the time evolution of states according to Hamiltonian (\ref{eq:Htotal}). We provide a comparison of the numerical results of the bipartite entanglement entropy $S(t)$  against  exact diagonalization. 

A many-body quantum state can be described by the linear superposition of $2^N$  basis vectors 
\begin{equation}
|\psi\rangle =\sum_{s_i} c_{s_1 s_2 \dots s_N} |s_1 s_2 \dots s_N\rangle,
\label{qsA}
\end{equation}
where $|s_i\rangle$ are the eigenstates of $S^z_i$  with $|\uparrow\rangle$ or $|\downarrow\rangle$ spin states. The quantum state \ref{qsA} can be represented in the form of a matrix product state (MPS) as
\begin{eqnarray}
|\psi\rangle &=  \sum_{s_i}  u^1_{s_1l_1} u^2_{l_1s_2l_2} u^3_{l_2s_3l_3}\dots u^{N-1}_{l_{N-1}s_{N-1}l_N} u^N_{l_Ns_N}  \nonumber  \\
                  & \times |s_1 s_2 \dots s_N\rangle ,
\label{MPSA}
\end{eqnarray}
where $u^p_{ijl}$ are tensors of rank-$3$ at spin site $p$.  These are associated with the basis vectors generated by the tensor product $|s_{1}\rangle\otimes\cdots\otimes|s_{N}\rangle$ and can be represented by the diagrammatic notation shown in Fig.\ref{fig:A1}.  
\begin{center}
\begin{figure}[b]
\includegraphics[scale=0.34]{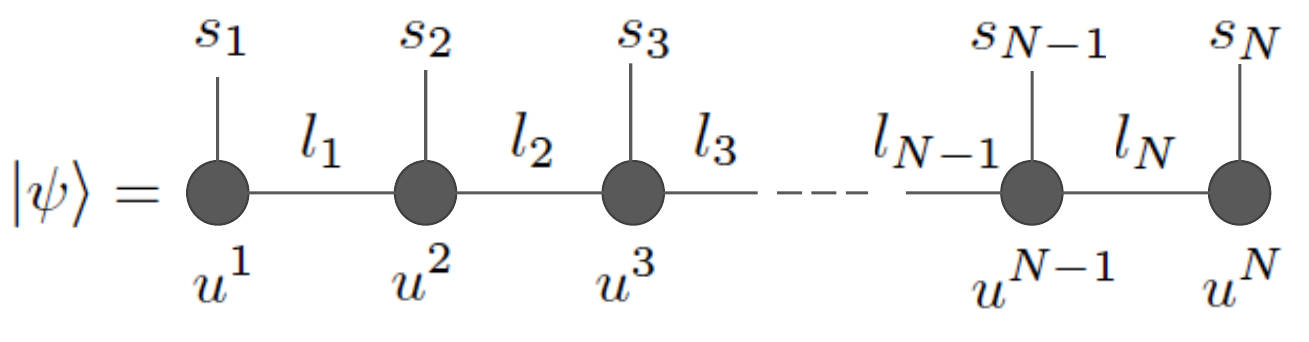}
\caption{\label{fig:A1} Diagrammatic notation of the Matrix Product State(MPS) tensor train.  }
\end{figure}
\par\end{center}
\begin{center}
\begin{figure}[t]
\includegraphics[scale=0.32]{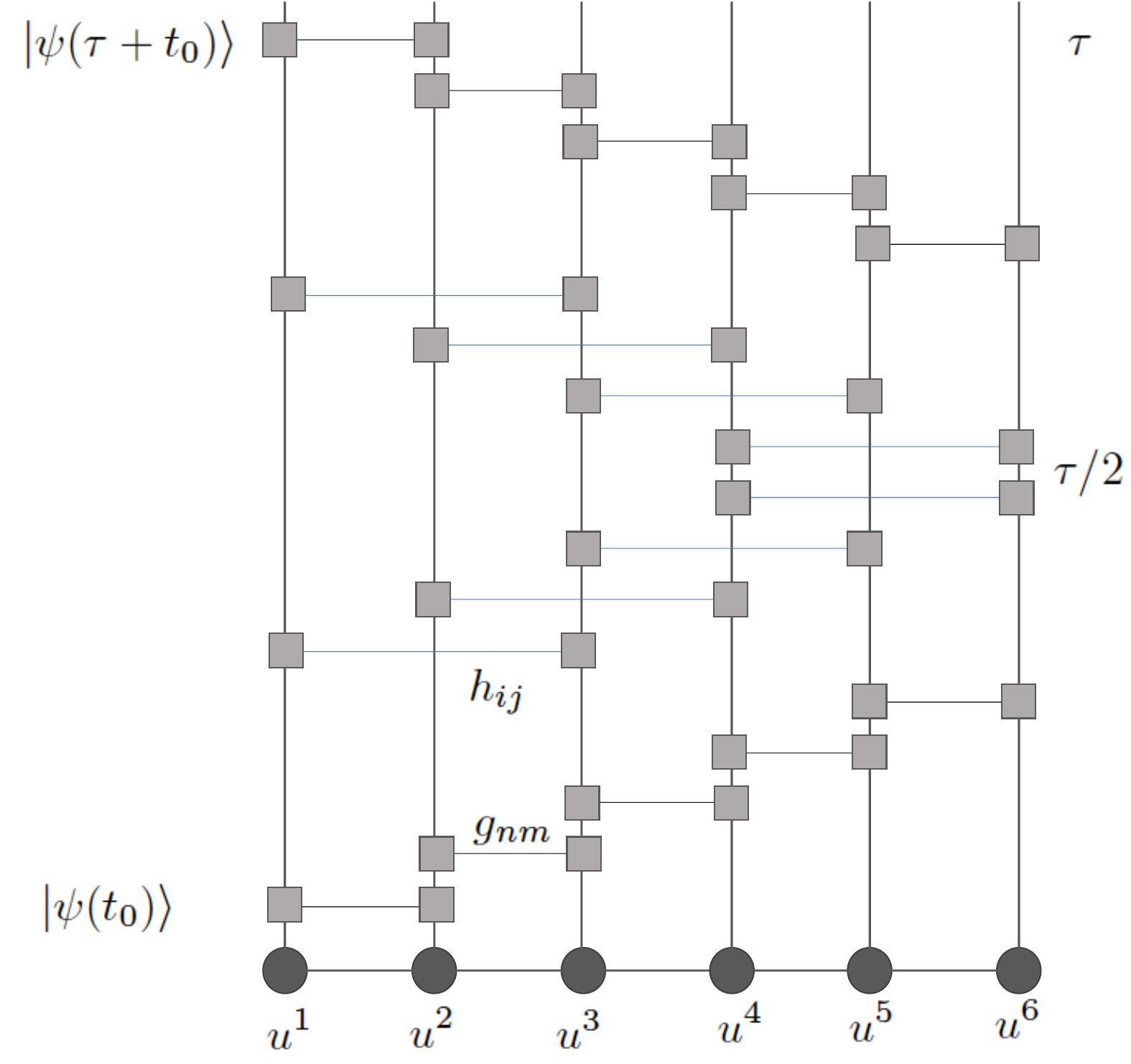}
\caption{\label{fig:A2} Diagrammatic notation of the time evolution algorithm (for time step $\tau$) following the second order Trotter decomposition. The algorithm is shown only for a system of $N=6$. The two rectangles connected by a thin horizontal line represent the operators $h_{ij}$ and $g_{nm}$. The solid circles represent the MPS at initial time $t_0$, $|\psi(t_0)\rangle$. }
\end{figure}
\par\end{center}
\begin{center}
\begin{figure}[b]
\includegraphics[scale=0.36]{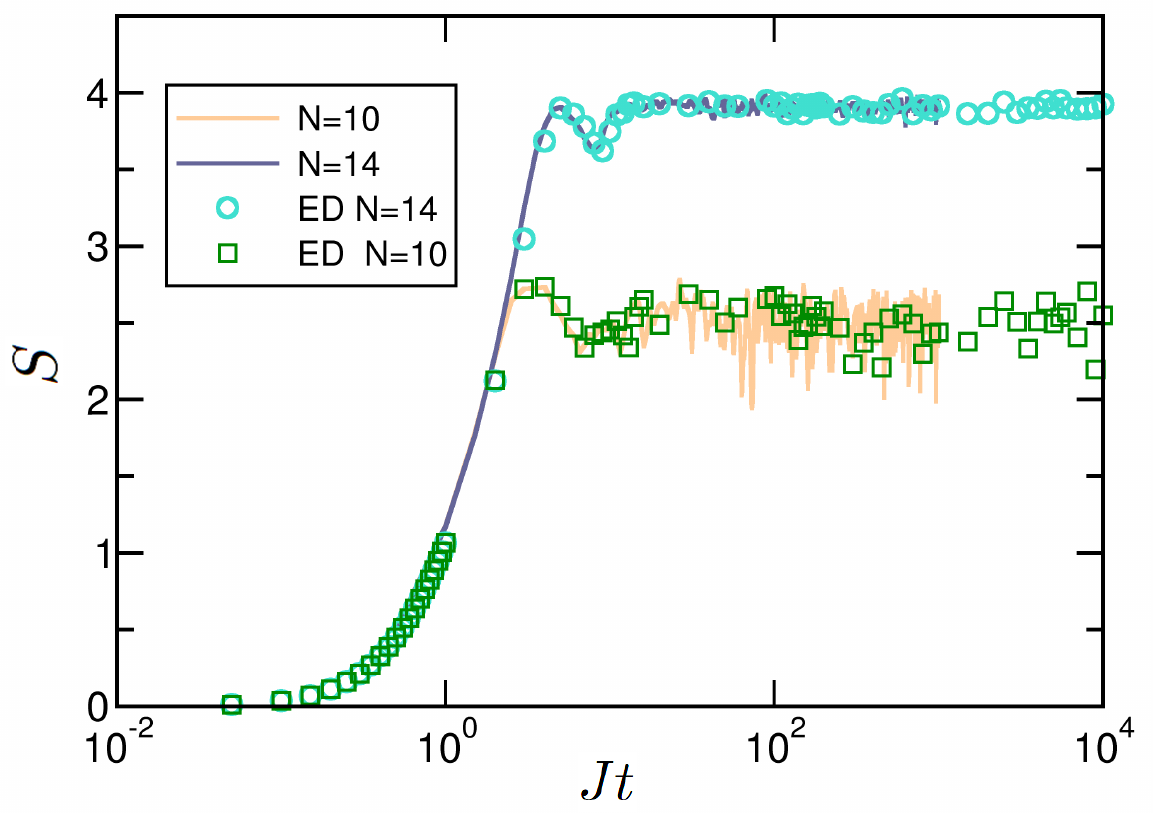}
\caption{\label{fig:A3} Plot of time evolution of the entanglement entropy of the system shown in Fig. \ref{fig:model} for $\gamma\delta=0.013$, $U^z_0/J=1$ and $U^{xy}_0/J=0$.  The solid lines represent the results of numerical calculation using the algorithm shown in Fig. \ref{fig:A2}.  The time step used is $\tau=0.05 J^{-1}$. A weight cutoff $10^{-7}$ is employed in {\it Julia} code truncating the MPS.  The open symbols represent the results from exact diagonalization (ED) for $N=10$ and $N=14$.  }
\end{figure}
\par\end{center}

The time evolution of a quantum state $|\psi\rangle$  is described by the unitary operator, $U(\tau)=e^{-i \mathcal{H}\tau}$, where $\mathcal{H}$ is the time independent Hamiltonian. The state at time $\tau+t_0$, $|\psi(\tau+t_0)\rangle$ can be given by applying the time evolution operator $U(\tau)$ to the initial state $|\psi(t_0)\rangle$
 \begin{equation}
 |\psi(\tau+t_0)\rangle=e^{-i \mathcal{H}\tau} |\psi(t_0)\rangle.
 \label{TimeEvol}
 \end{equation}
Hamiltonian (\ref{eq:Htotal}) can be written as the sum of locally interacting pairs of spins. We decompose the Hamiltonian into two parts: interactions within isotropic chains $\mathcal{H}_0=\mathcal{H}_A+\mathcal{H}_B \equiv \sum_{i,j} h_{ij} $, where the  pairs $i,j$ are interacting NN spins in the same chain; and  inter-chain interactions  $\mathcal{H}_{AB} \equiv \sum_{n,m} g_{nm}$, with the pairs $n,m$ denoting interacting NN spins in opposite chains. The locally interacting pairs of spins $h_{ij}$ and $g_{mn}$ are called gates.
The basis states are arranged in ascending order as shown in Fig. \ref{fig:model}. Thus, at low values of incommensuration $(\gamma\delta \ll 1)$, $j=i+2$ and $m=n+1$. 

The time evolution operator can be written as
\begin{equation}
U(\tau)=e^{-i \left(\mathcal{H}_{0}+\mathcal{H}_{AB}\right)\tau}.
\label{tEvoA}
\end{equation}
We now write  Eq. (\ref{tEvoA}) as a product of gates $h_{ij}$ and $g_{nm}$, so that the operator $U(\tau)$ can be contracted  with the MPS  in eq. (\ref{MPSA}) to numerically evaluate $|\psi(\tau+t_0)\rangle$.  Since $\left[\mathcal{H}_0,\mathcal{H}_{AB}\right]\neq 0$, we adopt a second order Trotter decomposition
\begin{equation}
\begin{split}
e^{-i \left(\mathcal{H}_{0}+\mathcal{H}_{AB}\right)\tau}\approx  e^{-i \mathcal{H}_{AB}\tau/2} e^{-i \mathcal{H}_{0}\tau/2}e^{-i \mathcal{H}_{0}\tau/2}e^{-i \mathcal{H}_{AB}\tau/2} \\ +O\left(\tau^3\right).
\end{split}
\label{eq:trotter}
\end{equation}
Similarly, we can expand the decomposition to individual gates. Equation (\ref{TimeEvol}) can be implemented numerically as shown in the diagrammatic  notation in Fig. \ref{fig:A2}.

We calculate the time evolution of the bipartite entanglement entropy with the numerical time evolution of the quantum state calculated using the algorithm represented in Fig. \ref{fig:A2}. 
We compare the time evolution of the entanglement entropy ($S$) calculated using the above numerical time integration procedure with the results from exact diagonalization (ED) in Fig. \ref{fig:A3} for $N=10$ and $N=14$. The two methods agree and give the same numerical results.

\end{document}